**Yizhuo Li, Lecture**
yizhuo_li@foxmail.com
**Chengdu University of Technology**
**Peng Zhou, Undergraduate**
zhou.peng@student.zy.cdut.edu.cn
**Chengdu University of Technology,**
**Fangyi Li, Undergraduate**
li.fangyi1@student.zy.cdut.edu.cn
**Chengdu University of Technology**
**Xiao Yang, Associate Professor**
310656749@qq.com
**Chengdu University of Technology**

# An Improved Reinforcement Learning Model Based on Sentiment Analysis

***Abstract:*** *With the development of artificial intelligence technology, quantitative trading systems represented by reinforcement learning have emerged in the stock trading market. The authors combined the deep Q network in reinforcement learning with the sentiment quantitative indicator ARBR to build a high-frequency stock trading model for the share market. To improve the performance of the model, the PCA algorithm is used to reduce the dimensionality feature vector while incorporating the influence of market sentiment on the long-short power into the spatial state of the trading model and uses the LSTM layer to replace the fully connected layer to solve the traditional DQN model due to limited empirical data storage. Through the use of cumulative income, Sharpe ratio to evaluate the performance of the model and the use of double moving averages and other strategies for comparison. The results show that the improved model proposed by authors is far superior to the comparison model in terms of income, achieving a maximum annualized rate of return of 54.5%, which is proven to be able to increase reinforcement learning performance significantly in stock trading.*
***Keywords:*** *Reinforcement learning, Deep recurrent network, Q learning, Investor sentiment, Sentiment quantitative indicator*

**JEL Classification: C51**

Yizhuo Li, Peng Zhou, Fangyi Li, Xiao Yang

## 1. Introduction

In recent years, with the widespread application of computer-related technologies, the financial industry has undergone tremendous changes on a traditional basis. Major Internet companies have invested in it to research innovative finance. At the same time, finding the most suitable quantitative trading strategy for the market has also become a hot topic of research by various institutions and scholars (Ta, Liu and Tadesse 2020). Based on the current development of the financial industry, this article found that even though the quantitative trading system has gradually begun to be used in the market, but the relevant algorithm strategies are not perfect(Liu et al. 2020). First of all, irrational investor behaviors is apparent even in Western stock markets such as mature US stocks, which means that investor sentiment has a significant impact on stocks (Li and Tang 2021). Because of this, many researchers have tried to construct various economic sentiment dictionaries in multiple languages through the dictionary to calculate the text sentiment of news media to quantify investor sentiment(Jiang, Meng and Tang 2021) . This has led to comparing the performance of various strategies to the implementation of competition language dictionaries. However, through literature research, this article finds that for listed companies in the market, the influence of media sentiment on the relative returns of individual stocks is asymmetry, which means that in the stock trading market, the degree of influence of positive and negative media sentiments on stock prices is not the same. And institutional investors mainly influence stock price changes through "buyer" behavior (Cai 2021). In short, changes in media sentiment may not directly affect changes in market stock prices, which are present in the stock markets of various countries (PAUL 2012). So this article assumes that using ARBR sentiment indicators combined with reinforcement learning methods to directly study the changes in current market long and short energy, this can achieve better returns in the stock market.

The ARBR sentiment indicator is also called the popularity indicator, which is composed of two indicators, AR and BR. The sentiment indicator ARBR believes that the stock market is engaged in a contest of long and short forces every day. Changes in stock prices are mainly caused by the contrast of long and short troops (Li 2018). The battle between the long and the short starts from a specific equilibrium price zone in the stock market. When the stock price is above this equilibrium zone, it indicates that the multi-party power is dominant; the short-term force is dominant when the stock price is below this equilibrium zone. Therefore, AR and BR indicators can more accurately measure the changes in the strength of the long and short sides in the market. This article uses the ARBR indicator to directly measure the changes in stock-related sentiment instead of building an economic sentiment dictionary.

In this article, our work is based on deep reinforcement learning-based stock market sentiment analysis and investment strategy algorithm research. First,

________________________________________________________________

we use the ARBR sentiment indicator to quantify the relationship between long and short forces behind stocks to obtain a dynamic coefficient and then use it as a tool to assist in-depth Cycle Q network. Secondly, based on the traditional DQN algorithm, in response to actual needs, we have improved the network structure and connection methods by including more stock technical indicators and reduced the dimensionality through the PCA algorithm and improved the profit and loss function of the network to explore the establishment of the complex state space of stocks is needed to learn to find a strategy that can maximize returns. Finally, the improved model also showed promising results in multiple sets of comparative experiments.

In summary, given the irrational behaviour of investors in the stock market, this article uses an improved algorithm of deep reinforcement learning based on the ARBR sentiment indicator to explore potential trading rules in the stock market. Compared with the traditional strategy of quantifying sentiment by constructing an economic sentiment dictionary, empirical evidence shows that the method proposed in the paper has achieved better benefits, and to a certain extent, broadened the application of reinforcement learning and sentiment analysis in financial markets. At the same time, a trading model that is more suitable for irrational investor markets also provides ideas for various investors to obtain better returns from their financial investments.

## 2. Literature review

More scholars have done related research in financial deep learning, which also provides ideas and help for the construction of deep neural networks in this paper. Scholars such as Zhang and Wang (2014) constructed a stock prediction model based on neural networks and solved the problem of high dimensionality of input data by means of genetic algorithm dimensionality reduction, and the results showed that genetic algorithm could improve the training efficiency of the model and enhance the prediction accuracy. Yang and Wang (2019) constructed a deep long and short-term memory neural network and applied it to the study of forecasting three different durations of more than 40 stock indices worldwide, and found that the LSTM neural network had strong generalization ability and was stable in predicting different durations of all indices. Sun and Bi (2018) selected convolutional neural networks and LSTM neural networks to construct up and down classification models respectively, based on which a high-frequency trading strategy was proposed and back-tested with the main asphalt futures contract as an example, and compared with the artificial neural network high-frequency trading strategy, the back-testing results showed that the high-frequency trading strategy based on convolutional neural networks and long and short-term memory neural networks had stronger profitability and better generalization ability is better. Wang and Xu (2019) used a variety of features extracted from stock data as input to an LSTM model, with the next day's rise as the output of the model for training. This resulted in a prediction of the next day's stock rise, which to some extent accounts for the regularity of stock returns.

Some scholars have done considerable research in reinforcement learning. Rong (2020) designed a method using deep reinforcement learning that uses time-series stock price data with the addition of news headlines for opinion mining, as well as a knowledge graph to exploit news about implicit relationships, and finally gives a summary. Zhu and Liu (2020) constructed a reinforcement learning framework with the Qlearning algorithm, using information on the rise and fall of a stock's opening and closing prices for several consecutive days as states, while the data for several consecutive days added predictive power to the model for guiding investments. Empirical analysis shows that this research approach to investment yields high and stable returns. Deng and Bao (2017) used deep neural networks to observe the dynamics of perceived financial markets and extract features, and used reinforcement learning for decision making to validate the robustness of the model in the stock and commodity futures markets. Bekiros (2010) proposed a reinforcement learning, complemented by an adaptive network fuzzy inference system, for high-frequency trading and achieved good results in real-world testing. Liu Quan et al. (2017) evaluated the performance of different strategy games under DRQN training and pointed out that DRQN outperformed the DQN model both in terms of learning curve and reward. Xiang (2018) used both univariate and bivariate methods to compare the payoff effects of DRQN models under different architectures and concluded that the LSTM architecture had better results in both cases. Uriel, Rolando and Ricardo (2020) proposed a cryptocurrency-based asset management model using deep Q and GRU networks and achieved better returns; Wei Liu (2021) proposed a deep Q network-based stock trading strategy model for the results of sentiment analysis of stock ticker data and commentary data making stock trading model more stable; Dai and Zhang (2021) demonstrated that the model with reinforcement learning approach outperformed the buy-and-hold strategy and MACD strategy in stock selection.

### 3. Methodology

### 3.1 Q-Learning algorithm

In a reinforcement learning problem, an object with perception and decision-making capabilities is called an Agent, and an Agent accomplishes a task by interacting with the external environment, which is the sum of the external environments that can be influenced by the actions of the intelligent body and give the corresponding feedback (Xiong et al. 2018). For an agent, it perceives the state of the environment and produces an action to make a decision; for the environment, it starts from an initial state and dynamically changes its state by accepting the action of the agent and giving a corresponding reward signal.

The Q-learning algorithm learning process can be described by the following: the Agent in the target environment can take the action space as $A$ and the state space as $S$, and make a transfer from the current state to the next state by means of a probabilistic transfer matrix $P$, while obtaining a reward $R$. The Q-value corresponding to the state to action is denoted by $Q(s,a)$.

We assume that the Agent's state at moment $t$ is $s_t$ and the action taken at this moment is $a_t$. The agent performs the action and moves to moment $t+1$, where the state changes to $s_{t+1}$ and the reward received is $r_t$. By updating the value of $Q(s,a)$ through all records of $(s_t, a_t, s_{t+1}, a_{t+1})$, it is possible to iterate continuously to find the optimal policy. The detailed formula can be expressed as follows.

$$\widehat{Q}(s,a) = \widehat{Q}(s,a) + \alpha \left( r + \gamma \max_a \widehat{Q}(s',a) - \widehat{Q}(s,a) \right) \tag{1}$$

In short, the Q-learning algorithm is a reinforcement learning algorithm based on Q-values, in which the variables of state and action are formed into a Q-table and the Q-values of the corresponding combinations are stored. The Q-value is used to continuously improve the strategy by constantly updating the actions that achieve the greatest reward, and finally giving an optimal solution. Its structure is illustrated in Figure 1.

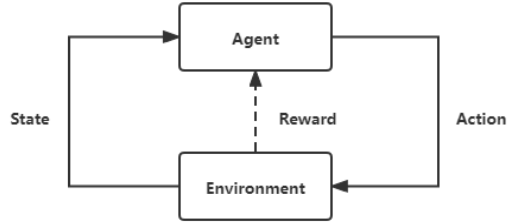

**Figure 1. Q learning model structure**

### 3.2 Deep Q Network

Minih et al. proposed a DRL algorithm for deep Q-networks by combining CNNs and Q-learning algorithms in traditional RL. This algorithm solves to some extent the problem of algorithm instability when using a nonlinear function approximator to represent a value function (Carta et al. 2021). DQN belongs to a kind of DRL, which is a combination of deep learning and Q-learning. When the combinations of states and actions are not exhaustible, the traditional Q-learning algorithm can no longer select the optimal action by looking up the Q-table, and a suboptimal solution that wirelessly approximates the optimal solution can be found by a deep Q-network without exhausting all combinations.

In a DQN neural network, the inputs are the state $s_1$ and the action space $\{a_1, a_2, \cdots, a_n\}$, and the outputs are the corresponding Q values for each action $q(s_1, a_1), q(s_1, a_2) \cdots q(s_n, a_n)$. Finally we choose the action corresponding to the largest Q value to perform the corresponding operation, and the Q value determination formula can be expressed as

$$Q(s_t, a_t) = R_{t+1} + \gamma \max_a Q(s_{t+1}, a) \tag{2}$$

The principle of DQN is to predict the Q estimate for each action by means of a neural network on the action space and select the action with the largest Q estimate to receive the corresponding reward. The structure is shown in Figure 2.

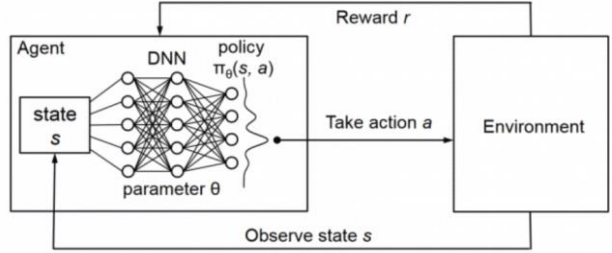

**Figure 2. Deep Q network model structure**

### 3.3 Quantification of market sentiment

In the introduction this paper points out that there is a correlation between large rises or plunges in the stock market and changes in public sentiment. However, as this effect has been shown to be asymmetric, this means that market sentiment affects stock rallies or plunges to different extents. We therefore use market trading data to directly reflect changes in long and short equity forces as an indirect reflection of changes in market sentiment.

The sentiment indicator ARBR consists of the popularity indicator AR and the willingness indicator BR. AR and BR are a pair of data reflecting the strength of long and short forces at different times through the analysis of historical stock prices, which can infer the current trading sentiment of the market and thus predict the reversal point of stock prices more accurately (Qiao and Yang 2016).

The AR indicator reflects the popularity of a target stock in a trading market by comparing the change in the opening, high and low prices over a given period, and it is based on the following formula.

$$AR = \frac{\sum_{i=1}^{N} \beta_i - \alpha_i}{\sum_{i=1}^{N} \alpha_i - \mu_i} * 100 \tag{3}$$

The BR indicator reflects the degree of willingness to buy or sell a target stock by comparing the position of the closing price in a given cycle to the price movement in that cycle, and it is expressed by the following formula.

$$BR = \frac{\sum_{i=1}^{N} \beta_i - \alpha_i^{'}}{\sum_{i=1}^{N} \alpha_i^{'} - \mu_i} * 100 \tag{4}$$

In the formulae for both indicators, where $\alpha_i$ is the opening price on day $i$,

$\beta_i$ is the highest price on day $i$, $\mu_i$ is the lowest price on day $i$, $\alpha_i'$ is the closing price of the previous day's trading on day $i$ and $N$ is the given time period.

### 3.4 PCA algorithm delas with feature vectors

In our research, the use of large data sets with multiple feature variables will undoubtedly bring more information to our study, but on the other hand, due to the increase in the size of the data set, the data collection and processing work becomes more difficult. In order to improve the performance of the model, we incorporate more stock technical indicators into the spatial state of the model, which to some extent improves the performance of the model, but also makes it significantly more difficult to process the data. This increases the complexity of the problem to a certain extent.

In this paper we use a 2-dimensional vector of z-score-processed price data and factor data for each technical indicator on a timeline as the state. Table 1 shows in detail how we store these states. This approach increases the level of abstraction of the current day's state while taking into account the influence of historical data on the current day. However, in the data, the factor data are mostly calculated from basic price data or volume turnover, and there is a strong correlation between the columns. Therefore, we use PCA to reduce the dimensionality of the data. By using PCA, we can transform a high-dimensional data set into a low-dimensional data set, removing and merging some unimportant features, and avoiding the impact of large stock technical indicator data on the efficiency of the model. At the same time PCA reduces the size of the analysed data while minimising the loss of data information, thus achieving a comprehensive analysis of the data (Chang and Wu 2015).

**Table 1. Storage mode of indicator factor**

| Date | Open | Close | … | Boll | … | Volume | MACD |
|---|---|---|---|---|---|---|---|
| t | $V_{(1,1)}$ | $V_{(1,2)}$ | … | $V_{(1,8)}$ | … | $V_{(1,j-1)}$ | $V_{(1,j)}$ |
| t-1 | $V_{(2,1)}$ | $V_{(2,2)}$ | … | $V_{(2,8)}$ | … | $V_{(2,j-1)}$ | $V_{(2,j)}$ |
| t-2 | $V_{(3,1)}$ | $V_{(3,2)}$ | … | $V_{(3,8)}$ | … | $V_{(3,j-1)}$ | $V_{(3,j)}$ |
| … | … | … | … | … | … | … | … |
| t-n | $V_{(n,1)}$ | $V_{(n,2)}$ | … | $V_{(n,8)}$ | … | $V_{(n,j-1)}$ | $V_{(n,j)}$ |

PCA is a frequently used method in the field of machine learning data mining, where the main purpose is to reduce the dimensionality of the data and retain the main information. The main idea of PCA is to map the original n-dimensional features onto another set of k-dimensional orthogonal bases. Because of the mapping process of the PCA algorithm, the features are no longer interpretable with each other after the PCA algorithm processing.

Yizhuo Li, Peng Zhou, Fangyi Li, Xiao Yang

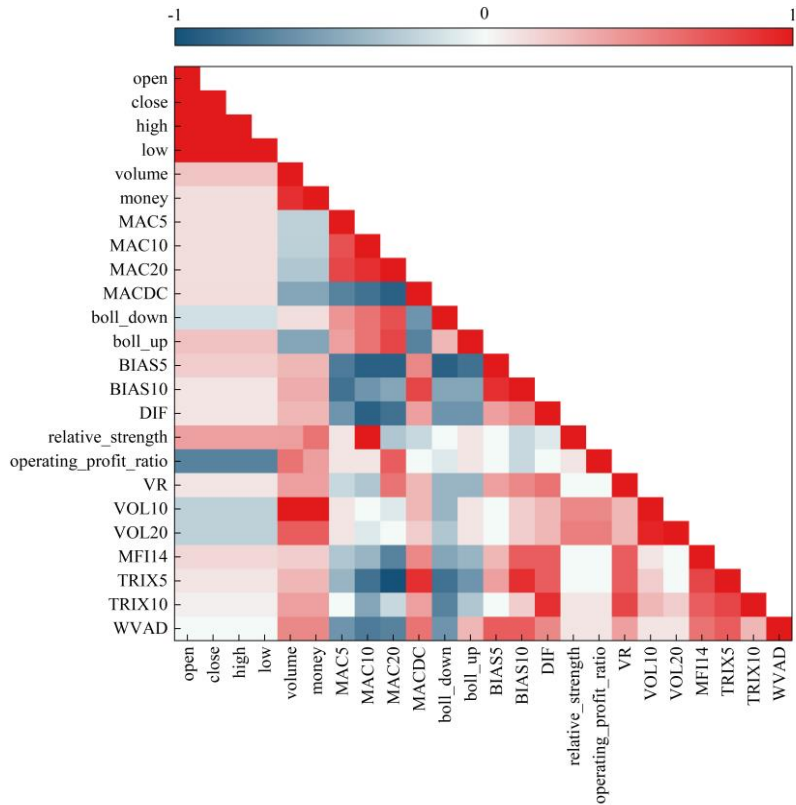

**Figure 3. Correlation charts for stock indicators**

The graph figure3 shows a heat map of the correlation between the various characteristic fields used, from which it can be seen that there is a large positive correlation between the volume and money fields and vol10 and vol20, and a large negative correlation between BIAS and MA. There is also a degree of correlation between the other variables. It is worth noting that after the PCA process each feature data is only correlated with itself, the other features are not correlated with each other two by two. Finally, after PCA processing we reduced the 24-dimensional feature vector data of the original data to 20-dimensions, which minimised the computational burden on the program while ensuring maximum retention of the original information.

### 3.5 Stock states and actions

In the construction of reinforcement learning models, the question of how to abstract the state space of the model is one of the central issues. Acting in the concrete financial domain, a state can be understood as a price position a stock is in. For stock trading, the most basic data describing a stock is the stock price. In addition to this, researchers have calculated some technical indicator factors based on some statistical knowledge, and some financial data of the company's operation can be used as the basis for state abstraction. This paper therefore uses stock price data and related technical indicators as an abstraction of the daily status of a stock.

In order to better represent the complex financial market, this paper encompasses as many relevant factors affecting stock buying and selling behaviors as possible as the spatial state of the stock. After several experiments, this study finally decided to use the grouped closing prices of stocks, 24 stock technical indicators after PCA dimensionality reduction, etc. as the state space of the target stocks, and its specific construction diagram is shown in Figure 4.

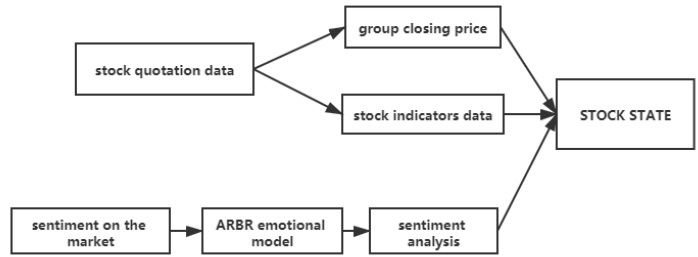

**Figure 4. Stock state structure**

The action space for reinforcement learning is the set of all valid actions of the model, which determines the range of actions used by the Agent. In this paper, we consider buying and selling of nine sector indices in the stock market, and the actions are only buy, sell and hold and watch, so this is the discrete action space.

The action space for reinforcement learning is the set of all valid actions of the model, which determines the range of actions used by the Agent. In this paper, we consider buying and selling of nine sector indices in the stock market, and the actions are only buy, sell and hold watch, so this is the discrete action space. We use the LSTM layer as the output layer, and the output is a one-dimensional tensor containing three elements, using an action selection strategy based on greedy rules. If the random number is smaller than a pre-defined greedy value, then the index value corresponding to the largest element in the output vector is selected as the action value, otherwise a random integer in the range -1 to 1 is selected as the action value. The action status is set as follows: when the action value is 1, it is defined as buy stock, when the action value is -1, it is defined as sell stock, when the action value is 0, it is defined as hold and wait, and the action space formula is expressed as follows.

$$a = \begin{cases} 1 \\ 0 \\ -1 \end{cases} \quad (5)$$

It is important to note that the execution of a sell operation means selling a specified amount of shares at the current time and paying a transaction fee. In this study we set the cost per trade to 0.1% of the trade amount and the model is assumed to operate in an ideal environment, which means that the historical price at the moment the order is placed is the actual price traded.

### 3.6 ARBR-DRQN Hybrid Model Development

In considering the structural design of the model, a modified ARBR-DRQN combination model is used in this study in order to allow the model to make better decisions from a large amount of data such as time series and technical indicators, as a way to deeply explore the hidden laws of the input information. The flow chart of the combined model is shown in Figure 5 and is roughly divided into three parts. The first part is to determine the timing of buying and selling stocks based on the ARBR sentiment indicator, the second part is to learn stock strategies based on the DRQN network, and the third part is to synthesize the action signals from the first two parts and execute the corresponding actions.

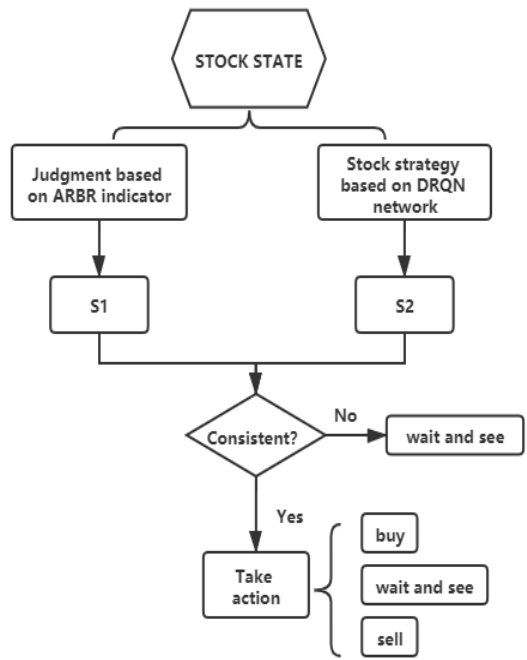

**Figure 5. Flow chart of deep Q model based on ARBR**

- **Stock Space:** Pre-processing of the basic stock ticker data by grouping, pre-processing of the daily technical indicator data and pre-processing of the daily ARBR values. The integrated all data is used as the spatial state of the stock.
- **Buy and sell strategies based on ARBR sentiment indicator:** Based on the spatial state of the stock, we analyse the stock based on ARBR values and construct a dynamic buy and sell point judgment model for the stock based on sentiment indicator. Through the dynamic comparison of the AR and BR values of the stock, we can determine the best operation at the current time, and issue the corresponding operation signal S1.
- **Stock trading strategy based on DRQN network:** A trading strategy model based on DRQN neural network is constructed through the spatial state of the stock. Using the model to get the buy and sell signal S2 and compares it with the model signal S1 in (2), and executes the operation if they are the same.

### 3.7 Reward function

The reward value refers to the reward the agent receives for each action performed. The trading model optimises the accuracy of the model and adjusts for subsequent actions by setting the reward value. Since this model studies high frequency stock data at the minute level, we define the maximum profit that can be obtained per minute after grouping as the reward value of the model.

$$r_t = P_t - P_{t-1} \tag{6}$$

In the formula, $r_t$ is the current reward value, $P_t$ is the price at time $t$ and $P_{t-1}$ is the price at the previous time. The reward value is the difference between the two momentary prices. When the current price is greater than the past price, the reward value is positive; when the current price is lower than the past price, the reward value is negative. Since the strategy of the model is to find the return under the best strategy, the final cumulative return can be expressed as follows:

$$R_t = \sum\nolimits_{t=1}^{T} r_t \tag{7}$$

## 4. Empirical Analysis of the Validity of ARBR-DRQN

This chapter conducts experimental validation on the CSI 9 sector indices to calculate the annualized returns and Sharpe ratios. An ablation experiment is also conducted to explore the experimental effects of the stock buying and selling judgment method incorporating the sentiment indicator ARBR and the stock buying and selling judgment method without ARBR, as well as to discuss the correctness of the buy and sell points of the model proposed in the paper. Besides, in order to explore the stability of the model, this chapter also sets up different initial investment amounts for validation and investigates the impact of different initial investment amounts on the annualized return and Sharpe ratio. Finally, we compare the deep Q-network trading strategy model based on the ARBR sentiment indicator with the double mean and buy-and-hold strategies to explore the magnitude between their annualized returns and Sharpe ratios.

### 4.1 Data Collection

Many quantitative trading systems for the Chinese equity market tend to ignore the impact of input data on the results, resulting in strategies whose back-testing results are far from the true application results. As the Chinese equity trading market is generally immature and investors are predominantly young and irrational, the validity and continuity of a strategy that achieves high returns on individual stocks or over a short period of time should be questioned in this context. To minimize the impact of the particular market environment on quantitative strategies, this study selects five years of one-minute stock data for the CSI Nine Sector Index from 4 January 2016 to 1 October 2021, as shown in table 2. This data includes the daily opening and closing prices of stocks, technical indicators and various other information subsequently used to process ARBR sentiment quantification.

We use grouping to pre-process the huge amount of data. We divide the minute data into 30-minute groups, with the first and last data of each group being the opening and closing prices of the group respectively. Since the closing price of the first group can be regarded as the opening price of the second group, it is only necessary to record the closing price of each 30-minute period in order to show the state of the stock at any given time in the empirical process.

At the same time, we calculate the log returns of the closing prices of 8 adjacent groups of stocks and treat the results as 8 features of the spatial state of the stocks in the current group at time. The features of these states will be fed back to the LSTM network for further processing by z-score normalization. Where $x$ is the original data, $\mu$ is the mean data and $\sigma$ is the standard deviation.

$$z = \frac{x - \mu}{\sigma} \tag{8}$$

**Table 2. Selected stock industry index**

| Index | Abbreviation | Code |
|---|---|---|
| CSI Materials Index | Materials | 399929 |
| CSI Telecommunication Services Index | Telecom | 399936 |
| CSI Utilities Index | Utilities | 399937 |
| CSI Industrials Index | Industrials | 399930 |
| CSI Financials Index | Financials | 399934 |
| CSI Energy Index | Energy | 399928 |
| CSI Consumer Index | Cons | 399932 |
| CSI Information Technology Index | IT | 399935 |
| CSI Health Care Index | Health Care | 399933 |

**4.2 Ablation Study**

After back testing, we count the returns of the model under different strategies. We compare the method incorporating the sentiment indicator ARBR with the method without ARBR, and the annualized rates as well as Sharpe ratios obtained are shown in Table 3. From Table 3, it can be seen that the improved stock trading strategy model based on deep Q-networks proposed in this chapter is effective and profitable, and is able to obtain a certain annualized rate of return on investment. Among them, the new strategy proposed in the thesis obtained the highest annualized return of 54.5% in the IT sector. It also achieved an annualized return of 10.08% even in the worst-performing Utilities sector. It is worth noting that the approach incorporating the sentiment-based ARBR has higher average annualized returns and higher average Sharpe ratio than the approach without the sentiment-based ARBR. Overall, the annualized returns and Sharpe ratios of the method incorporating the sentiment-based ARBR are higher than those of the method without. In addition, when ARBR is not fused, the strategy has a negative Sharpe ratio on the material index. This paper therefore concludes that fusing the two trading strategies can make the overall trading strategy model more stable, reduce the risk of misspecification and improve the model's returns.

Table 3. Ablation study results based on ARBR deep Q model

| Index | Method With ARBR | | Method without ARBR | |
|---|---|---|---|---|
| | Annual Return | Sharpe Ratio | Annual Return | Sharpe Ratio |
| Materials | 12.09 | 3.03 | 1.62 | -0.46 |
| Telecom | 43.73 | 13.58 | 20.42 | 5.81 |
| Utilities | 10.08 | 2.36 | 7.32 | 1.44 |
| Industrials | 31.56 | 9.52 | 21.36 | 6.12 |
| Financials | 49.95 | 15.65 | 41.2 | 12.73 |
| Energy | 19.47 | 5.49 | 13.17 | 3.39 |
| Cons | 16.09 | 4.36 | 15.02 | 4.01 |
| IT | 54.5 | 17.17 | 45.24 | 14.08 |
| Health Care | 14.61 | 3.87 | 14.08 | 3.69 |
| Average | 28.01 | 8.54 | 19.94 | 5.65 |

**4.3 Stability test**

In order to test the stability of the improved model returns, we explore the effect of different initial capital on the final stock returns. We vary the initial investment capital to RMB 100,000, RMB 200,000 and RMB 300,000 to test the index again, and the results are shown in the table 4. These three different amounts of initial capital were chosen as variables because the paper considers that these three gradations of capital settings are appropriate given the limited amount of capital that most individual investors have access to the stock market. As can be seen from the table 4, the change in initial capital does not have a significant impact on the annualized returns and Sharpe ratios of the stocks. This indicates that the model is successful in achieving relatively stable returns under different capital allocations

**Table 4. Results of returns at different initial amounts**

| Index | 100000 | | 200000 | | 300000 | |
|---|---|---|---|---|---|---|
| | Annual Return | Sharpe Ratio | Annual Return | Sharpe Ratio | Annual Return | Sharpe Ratio |
| Materials | 12.09 | 3.03 | 14.02 | 3.67 | 9.51 | 2.17 |
| Telecom | 43.73 | 13.58 | 40.27 | 12.42 | 38.79 | 11.92 |
| Utilities | 10.08 | 2.36 | 11.46 | 2.82 | 12.13 | 3.04 |
| Industrials | 31.56 | 9.52 | 33.38 | 10.13 | 36.96 | 11.32 |
| Financials | 49.95 | 15.65 | 40.29 | 12.43 | 55.43 | 17.48 |
| Energy | 19.47 | 5.49 | 22.45 | 6.48 | 16.29 | 4.43 |
| Cons | 16.09 | 4.36 | 15.31 | 4.10 | 17.11 | 4.70 |
| IT | 54.5 | 17.17 | 56.28 | 17.76 | 64.3 | 20.43 |
| HealthCare | 14.61 | 3.87 | 18.62 | 5.21 | 12.80 | 3.27 |

**4.4 Comparison of different strategies**

We compare the approach proposed in the paper with various strategy models, and the table5 shows the results of the comparison experiments of the stock trading strategy models. The comparison study selects the double moving average strategy, buy-and-hold strategy and the ARBR-DRQN proposed in this paper for comparison. From the table5, it can be seen that the stock trading strategy based on the ARBR deep Q network proposed in this paper outperforms both the double moving average strategy and the buy-and-hold strategy in terms of average annualised returns and Sharpe ratios. This also demonstrates that the ARBR-DRQN model is more stable in terms of returns, showing higher investment returns overall.

The least stable strategy is the double moving average strategy, which can achieve higher investment returns for certain stocks, but also has significant losses. Buy-and-hold, a typical passive investment strategy, offers better returns than the double moving average strategy.

Finally, we selected the relatively better payoff buy and hold strategy as the baseline to compare with the improved ARBR-based Q-neural network strategy in terms of payoffs. As shown in Figure 6, it can be found that the thesis strategy achieves higher returns overall.

**Table 5. Comparison of profit results under different strategies**

| Index | Evaluation Index | Hold and Buy | Double Moving Average | DRQN-ARBR |
|---|---|---|---|---|
| Materials | Annual Return | 1.42 | 2.1 | 12.09 |
| | Sharpe Ratio | -0.53 | -0.53 | 3.03 |
| Telecom | Annual Return | 4.82 | 15.33 | 43.73 |
|  | Sharpe Ratio | 0.61 | 4.11 | 13.58 |
| Utilities | Annual Return | -3.21 | -10.13 | 10.08 |
|  | Sharpe Ratio | NA | NA | 2.36 |
| Industrials | Annual Return | -3.17 | -4.59 | 31.56 |
|  | Sharpe Ratio | NA | NA | 9.52 |
| Financials | Annual Return | 3.29 | -1.15 | 49.95 |
|  | Sharpe Ratio | 0.1 | NA | 15.65 |
| Energy | Annual Return | 6.55 | -7.31 | 19.47 |
|  | Sharpe Ratio | 1.18 | NA | 5.49 |
| Cons | Annual Return | 29.51 | 9.42 | 16.09 |
|  | Sharpe Ratio | 8.84 | 2.14 | 4.36 |
| IT | Annual Return | 6.84 | -16.64 | 54.5 |
|  | Sharpe Ratio | 1.28 | NA | 17.17 |
| HealthCare | Annual Return | 12.25 | 12.94 | 14.61 |
|  | Sharpe Ratio | 3.08 | 3.31 | 3.87 |

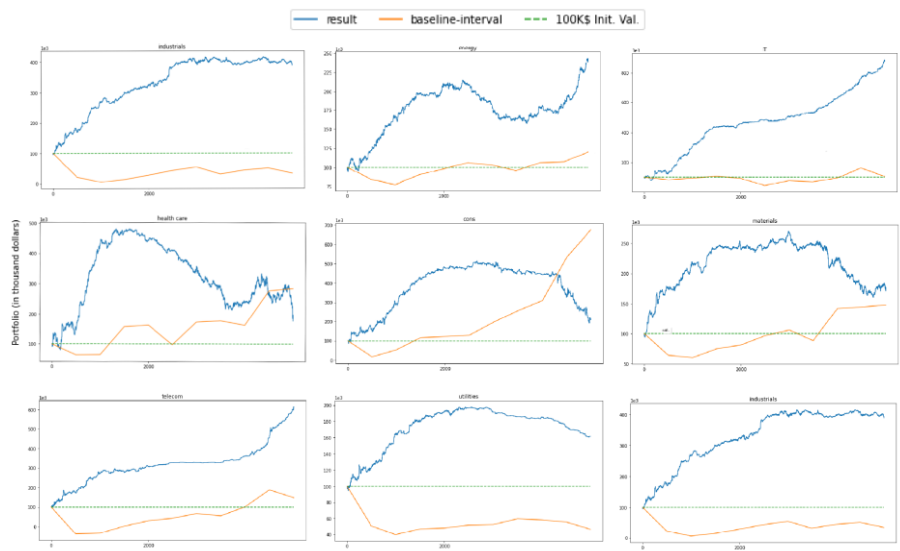

**Figure 6. Comparison of model and base-line revenue**

### 4.5 Buying and Selling point accuracy test

To further test the accuracy of the deep Q neural network based on ARBR for bid/ask point determination, in this paper, we calculate the number of buy and sell order executions as well as the correct rate of execution for each sector separately, using January 1, 2017 to December 31, 2017 as an example, as shown in the table 6. It is worth noting that execution trades are measured in days and that an order is considered to have successfully captured the timing of a trade if the share price rises after a buy or falls after a sell. As can be seen from the table, throughout 2017, even though there were occasional errors in judgement, overall the trading strategy model proposed in this paper was able to better guide the buying of shares at their lows and the selling of shares at their highs, which was experimentally to yield good returns

**Table 6. Accuracy of ARBR-DRQN model**

| ARBR assisted trading within 1000 days (trade in days) | | | | | |
|---|---|---|---|---|---|
| Index | Order | Order Placed | | Orders Correct | |
| Materials | Buy | 91 | 59.48% | 53 | 58.24% |
| | Sell | 62 | 40.52% | 28 | 45.16% |
| Telecom | Buy | 72 | 63.72% | 41 | 56.94% |
| | Sell | 41 | 36.28% | 26 | 63.41% |
| Utilities | Buy | 84 | 53.50% | 64 | 76.19% |
| | Sell | 73 | 46.50% | 52 | 71.23% |

| ARBR assisted trading within 1000 days (trade in days) |||||
| --- | --- | --- | --- | --- | --- |
| Index | Order | Order Placed || Orders Correct ||
| Industrials | Buy | 90 | 56.96% | 62 | 68.89% |
| | Sell | 68 | 43.04% | 42 | 61.76% |
| Financials | Buy | 77 | 33.48% | 52 | 67.53% |
| | Sell | 153 | 66.52% | 81 | 52.94% |
| Energy | Buy | 101 | 59.41% | 73 | 72.28% |
| | Sell | 69 | 40.59% | 56 | 76.81% |
| Cons | Buy | 84 | 60.00% | 66 | 78.57% |
| | Sell | 56 | 40.00% | 38 | 67.86% |
| IT | Buy | 94 | 58.75% | 67 | 71.28% |
| | Sell | 66 | 42.25% | 46 | 69.70% |
| Health Care | Buy | 63 | 57.80% | 47 | 74.60% |
| | Sell | 46 | 42.20% | 22 | 47.83% |

### 4.6 Kupiec test

In order to more accurately illustrate the effectiveness of the ARBR-DRQN network for stock purchase and sale timing, this paper further employs the likelihood ratio test proposed by Kupiec (1995) for statistical testing. Let $T$ be the total number of buy and sell orders issued by the ARBR-DRQN model, $F$ be the number of incorrect identifications by the network model, and $T-F$ be the number of times the buy and sell signals are correctly identified by the model. As a result, the statistic LR follows a $\chi^2$ distribution with degree of freedom 1, as shown in the following equation.

$$LR = -2\ln[(1-\alpha)^{T-F}\alpha^F] + 2\ln[(1-F/T)^{T-F}(F/T)^F] \qquad (9)$$

We have selected the period from 2016 to 2021 as the reference period, which roughly encompasses the complete market sentiment of bull, bear and shock markets and is more representative. In terms of the selection of moving periods, for the reliability of the results, the 12 periods shown in the table 7 below are selected in this paper.

At a given significance level $\alpha$, if the accompanying probability of the statistic LR is less than the critical value of the distribution, then the original hypothesis that the ARBR-DRQN is not very accurate in choosing the right time to buy and sell stocks will be accepted; conversely, if the accompanying probability of the statistic LR is greater than the critical value of the distribution, then the original hypothesis cannot be rejected and the ARBR-DRQN network's acuracy can be considered to be high. In this paper, the statistic LR is calculated at the 5% significance level and is shown in the table 7. The minimum value of the statistic LR is 0.095, and according to the study of Wang Chunfeng et al, the size of the statistic LR has a positive correlation with its concomitant probability, which is 0.230 when the LR value is 0.085. Since the LR of all the statistics in the table is greater than 0.085, the concomitant probability of all the statistics in the table will be greater than 0.230, which in turn is greater than the concomitant probability at the Since all the statistics in the table have a LR greater than 0.0.230, all the statistics in the table will have a probability of concomitance greater than 0.230 and thus greater than the threshold value for a 5% significance level distribution $\chi^2(1)$ . This indicates that the ARBR-DRQN network passes the statistical test for stock purchase and sale timing.

**Table 7. LR statistics for failure frequency tests**

| Period | 5 | 10 | 19 | 20 | 21 | 37 | 60 | 73 | 78 | 80 | 120 | 240 |
|---|---|---|---|---|---|---|---|---|---|---|---|---|
| Materials | 1.202679151 | 1.202679151 | 1.064029 | 0.778112 | 0.30684 | 0.997908 | 0.5737581 | 0.36375 | 0.4356329 | 0.773245 | 0.542804 | 0.529232 |
| Telecom | 1.271133974 | 1.271133974 | 1.10836 | 0.74264 | 0.17849 | 0.520523 | 0.8804609 | 0.98694 | 1.0210111 | 0.164799 | 0.126693 | 0.422475 |
| Utilities | 0.406951619 | 0.406951619 | 0.615726 | 0.842981 | 1.208092 | 1.011637 | 0.2512388 | 1.24891 | 0.1382878 | 0.528183 | 1.287119 | 0.208437 |
| Industrials | 0.69470835 | 0.69470835 | 0.71233 | 0.754749 | 0.48908 | 0.539925 | 0.4192871 | 0.65256 | 0.9661755 | 0.990257 | 0.767005 | 0.107898 |
| Financials | 0.753872978 | 0.753872978 | 0.975709 | 1.207443 | 0.787722 | 0.540526 | 0.7474884 | 0.57982 | 0.9935066 | 0.407147 | 1.003683 | 1.121222 |
| Energy | 0.554514999 | 0.554514999 | 0.978181 | 0.947281 | 0.927779 | 0.685749 | 0.9932533 | 0.58887 | 0.965804 | 0.456733 | 1.232499 | 0.566225 |
| Cons | 0.32532933 | 0.32532933 | 0.343003 | 0.295808 | 1.343794 | 0.746469 | 1.2993597 | 0.39935 | 1.1817397 | 0.2762 | 1.167173 | 0.270036 |
| IT | 0.594759861 | 0.594759861 | 0.956146 | 0.946809 | 0.145534 | 0.812977 | 0.3213393 | 0.91554 | 1.131969 | 0.367799 | 0.1437 | 0.16336 |
| Health Care | 0.464868898 | 0.464868898 | 0.747278 | 1.217503 | 0.76186 | 0.278797 | 0.6111825 | 0.92823 | 0.9747978 | 0.515061 | 1.163258 | 0.555702 |

### 4.7 Model Complexity

The computational complexity of a deep Q-network-based stock trading strategy model is related to the stock data dimension d, the number of stock data n and the model structure. The model structure includes the stock buying and selling point judgment method based on the sentiment indicator ARBR and the deep Q network. We note that the complexity of the stock buying and selling point judgment model based on ARBR is t, the complexity of the model of the deep Q network is q, and the structure of the two models in the stock trading strategy model based on the deep Q network is a parallel structure with no loops, then the complexity of the model structure of the stock trading strategy based on the deep Q network is $O(t+q)$. With a stock data dimension of d and a stock data volume of n, the computational complexity of the whole model is $O(nd(t+q))$, which is much lower than that of the sentiment stock trading model built through financial dictionaries or big data analytics.

### 5. Discussion

This study provides a stock trading strategy model based on the deep Q network of the sentiment indicator ARBR after doing research. The diversity of stock data is taken into account while creating the stock state space, and additional stock indicators and market data are combined to fully investigate the stock state. For the design of stock trading strategy models, a space representation is employed. Simultaneously, we present a stock trading point judgment technique based on ARBR and a framework of deep Q network integration to build a stock trading strategy that increases the strategy's stability and lowers investment risks. The experimental findings reveal that the model provided in this research outperforms the double moving average strategy and the Hold and Buy strategy on the two evaluation indicators of return rate and Sharpe ratio when tested on China Securities' 9 key industry indices. Furthermore, the integrated framework's trading strategy provides greater returns and stability than a single trading strategy model. As a result, the model described in this thesis is distinct and superior to existing tactics, and it has significant research and practical application value.

## 6. Fund

This research was supported by the Project of Ministry of Education of the People's Republic of China (202002121038) and the Young Core Instructor Development Program of Chengdu University of Technology.

___________________________________________________________________